\documentclass[twocolumn,showpacs,preprintnumbers,amsmath,amssymb]{revtex4}
\usepackage{graphicx}
\usepackage{dcolumn}
\usepackage{bm}

\newcommand \beq{\begin{eqnarray}}
\newcommand \eeq{\end{eqnarray}}
\newcommand \bea{\begin{eqnarray}}
\newcommand \eea{\end{eqnarray}}

\newcommand \bq{{\mathbf q}}

\usepackage{amsmath}
\usepackage{amssymb}

\def\psekibun{\int \frac{\mathrm{d}^3p}{(2\pi )^3}}

\def\ln {\mbox{ln}}
\def\tr {\mbox{tr}}
\def\exp {\mbox{exp}}

\def\bpi {\mbox{\boldmath{$\pi$}}}
\def\btau {\mbox{\boldmath{$\tau$}}}

\def\pq2 {((p+\frac{q}{2})^2-M_0^2)((p-\frac{q}{2})^2-M_0^2)}
\def\Ep {E_p}

\def\bq {\mbox{\boldmath{q}}}

\def\simle{\mathrel{
     \rlap{\raise 0.511ex \hbox{$<$}}{\lower 0.511ex \hbox{$\sim$}}}}

\begin{document}
\preprint{UT-Komaba/12-3}
\title{Equation of state of a meson gas  \\ from 
the PNJL model for interacting quarks}
\author{
Kanako Yamazaki\footnote{GCOE Research Assistant at the Physics Department of the 
Graduate School of Science, University of Tokyo, Hongo; e-mail: kyamazaki@nt1.c.u-tokyo.ac.jp }  
and T. Matsui\footnote{e-mail: tmatsui@hep1.c.u-tokyo.ac.jp}}
\affiliation{Institute of Physics, University of Tokyo, Komaba,Tokyo, Japan}

\vskip 20pt

\begin{abstract}

We compute the equation of state of hadronic matter at finite temperature with zero 
net baryon density by the Nambu-Jona-Lasinio model for interacting quarks in uniform 
background temporal color gauge fields.  
In the low temperature confining phase, unphysical thermal quark-antiquark excitations 
which would appear in the mean field approximation, are eliminated by enforcing 
vanishing expectation value of the "Polyakov-loop" of the background gauge field, 
while in the high temperature confining phase this expectation value is taken as close to 
unity allowing thermal excitations of free quarks and antiquarks. 
Mesonic excitations in the low temperature phase appear in the correlation energy 
as contributions of collective excitations.  
We describe them in terms of thermal fluctuations of auxiliary fields in one-loop (Gaussian) 
approximation, where pions appear as Nambu-Goldstone modes associated with dynamical 
symmetry breaking of the chiral symmetry in the limit of vanishing bare quark masses.
We show that the equations of state reduces to that of free meson gas with numerically small 
corrections arising from the composite nature of mesons.

\end{abstract}
\pacs{11.10.Wx,11.30.Rd, 12.38.Mh,25.75.Nq}
\maketitle

The Nambu-Jona-Lasinio (NJL) model, originally introduced to describe profound consequences of 
presumed underlying chiral symmetry in hadron dynamics in analogy to the BCS theory of 
superconductivity\cite{NJL61}, 
has been widely used to describe effective quark dynamics respecting the symmetry\cite{HK94}.   
When it is applied in the mean field approximation to a finite temperature system, however, it 
generates thermal excitations of quark quasiparticles even at low temperatures, 
where quarks are to be confined in individual mesons and baryons.
 
A phenomenological model which removes these unphysical modes of excitations 
was proposed by Fukushima\cite{Fuk04} by making use of the Polyakov loop 
which is known to characterize confinement-deconfinement transition 
in strong coupling pure gauge theories.\cite{Pol78}
This procedure eliminate the quark single particle distribution by phase cancellations, 
leaving behind excitations of triad of quark quasiparticles in color singlet configuration. 
The quark-liberating deconfining transition was described as a change of the expectation value of 
the Polyakov loop treated as an order parameter with effective Landau-Ginzburg type free energy functional.   
Fukushima's model has been reformulated as a mean field theory in a uniform background temporal gauge field 
and has been studied extensively by other groups \cite{Megias:2004hj,RTW06, RRW07,RHRW08,Blaschke:2007np}.
Their model, referred to as the PNJL model, allows to incorporate mesonic correlations beyond
mean field approximation\cite{RHRW08,Blaschke:2007np,HABMNR07}, as had been studied previously 
in the original NJL model\cite{HKZ94,FB96}, 
in analogy to the classic example of plasma oscillation.\cite{GB57}
Numerical computation with the model has been shown to reproduce the equation of 
state calculated by the Monte Carlo simulation of the lattice gauge theory at small non-zero 
chemical potential rather well. 
The model has been used to compute susceptibilities of hadronic matter at finite chemical potential\cite{SFR07}, 
studied in the regime of complex chemical potential\cite{SKKY08} and very recently 
used to study the effect of confinement in the study of QCD phase diagram\cite{PB12}.

It is the purpose of these notes to re-examine the model at zero chemical potential and show 
that equation of state of a meson gas can be derived explicitly at low temperatures by the method 
of auxiliary fields which physically express effective meson fields build-up as a quark-antiquark 
bound states. 
Similar results have been obtained earlier by others by numerical computations of the 
mesonic correlation energies with slightly different models\cite{RRW07,RHRW08}, but we will 
show explicitly that in the chiral limit the equation of state becomes that of a gas of mesons and 
a correction due to the composite nature of mesons as the individual excitations of the quark triad.  
It is found that the former dominates the equation of state at low temperatures and the latter do not 
alter the free meson gas result significantly. 
As the temperature increases the mean field contribution from underlying quark quasiparticle 
excitations become important due to the non-vanishing Polyakov loop expectation value and will 
eventually dominate the free energy 
in deconfining phase.  
Collective mesonic excitations melt into the continuum of quark anti-quark excitations. 

We begin with an Euclidean path-integral expression of the partition function for the NJL model 
of quarks in external temporal color gauge field:  
\begin{equation}
Z ( T, A_4 ) = \int [d q ][d \bar q ]  \exp{ \left[  \int_\beta d^4 x 
{\cal L}_{\rm NJL} ( q, {\bar q}, A_4 ) \right] }
\end{equation}
where we have used an abbreviate notation for integral over the space and imaginary time ($d\tau d^3 r =d^4 x$),  
from $-\infty$ to $\infty$ for each of three space coordinates and from 0 to $\beta = 1/T $ for imaginary time $\tau$,  
and the effective Lagrangian is given by 
\begin{equation}
{\cal L}_{\rm NJL} ( q, {\bar q}, A_4 ) = {\bar q} ( i \gamma^\mu D_\mu  - m_0 ) q 
 + G \left[ ( {\bar q} q )^2 + (i{\bar q} \gamma_5 \tau q)^2 \right]  
\end{equation}
for two flavor light quarks, $ {\bar q} = ( {\bar u}, {\bar d} ) $ in external (classical) temporal color gauge field, 
$D_\mu = \partial_\mu + g A_0  \delta_{\mu, 0} $ and $m_0$ is the bare quark mass which breaks the chiral 
symmetry explicitly.  Here we have used a standard Minkovsky metric notation with the real time 
being replaced by the imaginary time $\tau = i x_0$ and the time component of the gauge field replaced by $A_4= i A_0$
\cite{KG06}.
We note here that quark fields are the only dynamical variables which describes thermal excitations of the system, 
while the temporal component of SU(3) gauge fields $A_4 = \frac{1}{2} \lambda^a {\cal A}^a_4$,  
$\lambda^a$ being the 3 $\times$ 3 Gell-Mann matrices, only plays a side role 
imposing constraints on color configurations of thermal quark excitations.  
In the following calculation, we take Ansatz of diagonal representations for $A_4$ 
along with the earlier works\cite{Fuk04,RTW06, RRW07,RHRW08}.

To describe mesonic excitations, we introduce four auxiliary bosonic fields $\phi_i = (\sigma, \bpi) $ coupled 
to quark densities $({\bar q} q, i {\bar q} \gamma_5 \btau q)$ by multiplying $Z (T, A_4) $ by a constant Gaussian integral. This procedure, known as the Hubbard-Stratonovich transformation, converts the four point NJL quark interaction to Yukawa coupling 
of the bosonic fields $\phi_i$ to corresponding four quark densities.  
Performing the Grassmann integral over the quark Dirac fields, the partition function becomes
\begin{equation}
Z ( T, A_4 ) = \int [d \phi ] e^{ - I ( \phi,  A_4 ) } \label{PF}
\end{equation}
where the exponent is given by the integral 
\begin{eqnarray}
I ( \phi,  A_4 )  & = &  \frac{1}{4G} \int_\beta d^4 x  \left( (\sigma - m_0)^2 + \bpi^2 \right)
\nonumber \\
& & \quad  -  \rm{Tr}  \ln \left[ \beta \left( i \gamma^\mu D_\mu  + \sigma + i   \gamma_5 \btau \cdot \bpi \right)   \right]  
\label{Lmeson}
\end{eqnarray}
where the trace in the second term is taken over the arguments of quark fields, including the space-time coordinates with anti-periodic boundary condition 
in the imaginary time axis, Dirac gamma matrices and the isospin and color indices.  

We evaluate the functional integral over the mesonic auxiliary fields by the method of steepest descent. Let $\phi _0(\sigma _0, \bpi_0)$ give local minimum value of integrand.   
Hereafter we choose $\bpi_0 = 0$, along with $\sigma_0 =  M_0$, which evidently satisfy stationary conditions for effective pion fields. Expanding the effective action in a power series of $\phi -\phi _0$ up to the quadratic term and performing the gaussian integral over the bosonic fields in eq.(\ref{PF}), we find 
\begin{equation}
\Omega (T,A_4) = T \left( I_0 +\frac{1}{2} {\rm Tr_M} \ln \frac{\delta I}{\delta \phi _i\delta \phi _j} \right)
\end{equation}
up to one-loop fluctuations of the collective meson fields, where the trace is to be performed over the 
internal coordinates of the auxiliary meson fields.

The thermodynamic potential in the mean field approximation, $\Omega_{\rm MF} (T,A_4)$, or the corresponding 
pressure $p_{\rm MF}  (T,A_4)$, is retrieved from the leading term $I_0$ by the relation: 
$\Omega_{\rm MF} (T,A_4) = T I_0  = - p_{MF} V$.
Converting the coordinate representation to the momentum representation and evaluating the discrete sum over the quark Matsubara frequencies by the standard 
method of contour integration\cite{FW71}, we find
\begin{eqnarray}
 p_{\rm MF} (T, A_4)  = 
2 N_f  \int \frac{d^3 p}{(2\pi)^3} \frac{p^2}{3 E_p}  \tr_c \left[ f ( E_p + i g A_4 ) \right.
\nonumber \\
  \qquad + \left. f ( E_p - i g A_4 ) \right]   + p_{\rm MF}^0  (M_0) 
\label{pmf}
\end{eqnarray}
where the first term is the pressure of the quark quasiparticle in the external gauge fields with distribution,
\begin{equation}
f ( E_p \pm i g A_4 ) = \frac{1}{e^{\beta ( E_p \pm i g A_4) }+1 } 
\label{qspdf}
\end{equation}
with energy $E_p = \sqrt{p^2 + M_0^2}$ and 
\begin{eqnarray}
p_{\rm MF}^0 (M_0) 
& =  &   3 \times 2 N_f  \int^{\Lambda}  \frac{d^3 p}{(2\pi)^3} E_p - \frac{1}{4G} (M_0 - m_0)^2 
\nonumber \\
\label{condpress}
\end{eqnarray}
is the pressure exerted by the quark quasiparticles in the "Dirac sea": It requires momentum cut-off 
$\Lambda $.
The factor 3 in the first term of  (\ref{condpress}) accounts for color and the factor 2 for spin degeneracies. 
Note that this "vacuum pressure" does not depend on the external background gauge potential due to the cancellation of the effects of the potential on particle and antiparticle. 
It depends on temperature indirectly through the temperature dependence of the effective quark mass $M_0$.  The vacuum pressure should be removed by the renormalization condition.
The quark pressure should also be removed in the low temperature phase where quarks are confined in hadrons.

In the mean field approximation, the value of the quark quasiparticle mass $M_0$ is determined by 
the stationary condition of $I_0$ with respect to $M_0$. This procedure leads to 
\begin{eqnarray}
M_0 - m_0 
& = & 8 G N_f  \tr_c \int^{\Lambda}  \frac{d^3 p}{(2\pi)^3} \frac{M_0}{E_p} \left( 1 - f ( E_p + i g A_4 ) \right.
\nonumber \\ 
& & \qquad \qquad -  \left. f ( E_p - i g A_4 ) \right) . \label{gap}
\end{eqnarray} 

In the above expressions, the constant temporal gauge field $A_4$ appears as a phase factor together 
with the quark quasiparticle energy in the single particle distribution function (\ref{qspdf}).  
It looks very similar to the gauge invariant (path-ordered) Polyakov loop phase integral, 
\begin{equation}
L ({\bf r} ) = {\cal P} \exp \left[   i g \int_0^\beta d \tau A_4 ({\bf r}, \tau )  \right] 
\end{equation}
whose thermal expectation value measures the extra free energy associated with the color charge 
in fundamental representation fixed at a spatial point $\bf r$. 
Although this connection is not strict,  
since quarks are moving in a uniform background gauge field not fluctuating either in space or in imaginary time, 
we adopt the procedure of Fukushima, along with the TUM group, to replace the phase factor 
in the quark quasiparticle distribution function by the thermal average of the Polyakov loop.

We replace 
$\langle \frac{1}{3} \tr_c  f ( E_p + i g A_4 )  \rangle $ by
\cite{HABMNR07}
\begin{equation}
 f_\Phi ( \Ep)   =
\frac{ \bar{\Phi } e^{2\beta \Ep } + 2\Phi e^{\beta \Ep }+ 1}
{e^{3\beta \Ep } +3\bar{\Phi } e^{2 \beta \Ep } + 3 \Phi e^{\beta \Ep }+ 1}
\label{averagedis}
\end{equation}
where $\Phi = \frac{1}{3}  \langle \tr_c L  \rangle$, and $\bar{\Phi} = \frac{1}{3}  \langle \tr_c L^\dagger  \rangle$.
The anti-quark distribution, $\bar{f}_\Phi ( \Ep ) \equiv  \langle \frac{1}{3}\tr_c f ( E_p - i g A_4 )  \rangle $, 
is obtained from the quark distribution by the replacement of $\Phi$ by $\bar{\Phi}$ and vice versa. 
At zero chemical potential, two distributions are identical and $\Phi$ is real so that $\Phi = \bar{\Phi}$.
Very interesting observation here is that although in the deconfining phase where $\Phi = \bar{\Phi} =1$, one would simply have ordinary quark distribution, 
\begin{equation}
\left.  f _\Phi ( \Ep) \right|_{\Phi =1}  = \frac{1}{ e^{\beta E_p}+1 } ,
\end{equation}
in the confining phase, where $\Phi = \bar{\Phi} = 0$, we have instead, 
\begin{equation}
\left.  f _\Phi ( \Ep) \right|_{\Phi =0} 
= \frac{1}{ e^{3\beta E_p}+1 } 
\end{equation}
which may be interpreted as a distribution function of triad of quark quasiparticles exciting together in color singlet configuration. 
It looks something like a baryon although no effect of interaction is taken into account between three quarks to
form a baryon, as reflected by the same value of momenta for all three quarks. 
We also note that the degrees of freedom of excitations is not reduced to 1/3 of color triplet free quark excitations. However, there still exist reduction of effective degrees of freedom due to the change of the phase space of quark momentum $p$ and that of quark triad momentum $3p$:
The "triad" carries the energy $E_{\rm tri} = 3E_p$, the momentum $p_{\rm  tri} = 3p$ and the mass 
$M_{\rm  tri} = 3M_0$.
This implies that the Lorentz invariant phase space integral over the quark momentum is reduced when written in term of the triad momentum.

Here we determine $\Phi$ phenomenologically by adding an effective potential $\mathcal{U}(T, \Phi )$:
$\Omega (T, \Phi ) = \langle \Omega (T, A_4) \rangle + V{\cal U} ( T,  \Phi ) $.
The parameters of ${\cal U} ( T,  \Phi )$ are chosen so that $\Phi =0$ for the low 
temperature confining regime where thermal quark quasi-particle excitations are forbidden,  
while $\Phi \simeq 1 $ above deconfining temperature, quite opposite to the normal behavior of 
the usual order parameter.
We note here that, 
although gluon fields are not treated as dynamical variables, gluon excitations may be included
in a phenomenological fashion through the effective potential by setting,
${\cal U} (T, \Phi=1) = - \frac{16 \pi^2}{90} T^4$
so that the equation of state matches to that of a free quark-gluon plasma at asymptotically high temperatures.

The equation of state in the mean-field approximation to the PNJL model is now given as a maximum of 
\begin{eqnarray}
p_{\rm MF} (T, \Phi) & = & 
4  \int \frac{d^3 p}{(2\pi)^3} \frac{p^2}{3 E_p}  f_\Phi ( E_p )  - {\cal U} (T, \Phi) 
\nonumber \\
& & \qquad +  p_{\rm MF}^0 (M_0) - \Delta p_{\rm vac}
\label{mfpressure}
\end{eqnarray}
with respect to variation of $\Phi$ and the quark quasiparticle mass $M_0$, which measures the magnitude 
of chiral condensate, is determined by 
\begin{equation}
M_0 - m_0 = 12 G N_f \int^{\Lambda}  
\frac{d^3 p}{(2\pi)^3} \frac{M_0}{E_p} \left( 1 - f_\Phi ( E_p ) - \bar{f}_\Phi ( E_p )  \right) .
\label{gap0}
\end{equation}
The constant $ \Delta p_{\rm vac}$ is chosen so that the pressure becomes zero at zero temperature and the cut-off 
at $\Lambda$ in the momentum integral in (\ref{gap0}) is applied only to the vacuum component.    

We show the temperature dependence of $\Phi$ and $M_0$ obtained from these mean-field equations in Fig. 1.  
In the chiral limit $m_0 = 0$, the gap equation (\ref{gap0}) possesses a non-trivial solution $M_0 \neq 0$ 
only at temperatures below a critical temperature $T_c$. 
In the ordinary NJL model, with no "Polyakov loop prescription", the chiral transition takes place at relatively low temperature,
while in the pure gauge lattice simulation, the confinement-deconfinement transition takes place at higher temperature 
as a first order transition with the discontinuous change in the value of $\Phi$.  
In the PNJL model, these two distinct transitions interfere each other through the quark quasiparticle loops.  
As a result, the two transitions takes place at a similar temperature:  the transition temperature for deconfinement becomes lower and 
becomes smooth crossover transition, while the chiral transition takes place at higher temperature, since the quark excitations are 
suppressed at low temperatures by small expectation value of the Polyakov loop.    
With a finite bare quark mass $m_0$, 
the chiral symmetry is broken explicitly and the chiral transition also become a smooth cross over transition. 

\begin{figure}[htbp]
\begin{center}
 \includegraphics[clip,width=70mm]{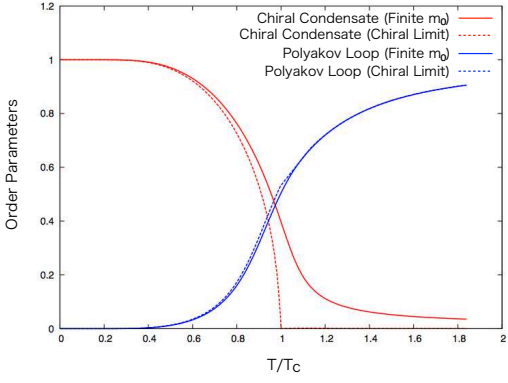}
\end{center}
\caption{
Temperature dependence of effective quark mass $M_0$ (red) and expectation value of the Polyakov loop (blue):
the dashed curves are for vanishing bare quark mass (the chiral limit) and the solid curves are with finite bare quark mass $m_0$ 
which is chosen to reproduce the pion mass $m_\pi = 140$MeV.  
}
\end{figure}

Now we come to our main task to evaluate mesonic correlation energy.    
The contribution of the mesonic correlation energy to free energy, $\Omega_M$, 
or the corresponding pressure $p_M = - \Omega_M/V$,  is given by
\begin{eqnarray}
p_M (T, A_4)  =
- \frac{T}{2} \sum_n \int \frac{d^3 q}{(2\pi)^3} \left\{ 
\ln \left[ \beta^2 \left( (2 G)^{-1} - \Pi_\sigma \right) \right]  \right. 
\nonumber \\
\qquad \qquad 
\left. + 3 \ln \left[ \beta^2 \left( (2 G)^{-1}  -  \Pi_\pi \right) \right]  \right\}\ \ \ \ 
\label{Mpressure}
\end{eqnarray}
where the sum is over the bosonic Matsubara frequencies $\omega_n = 2n \pi T$ which appear in the arguments of
meson self energies: 
\begin{eqnarray}
\Pi_{\sigma/\pi} (\omega_n, q,  A_4) & = & \Pi_{\sigma/\pi}^1 ( A_4 ) + \Pi_{\sigma/\pi}^2  (\omega_n, q,  A_4) 
\end{eqnarray}
The "tad pole" term 
\begin{equation}
\Pi_\sigma^1 =  \Pi_\pi^1 = - 2 T \sum_n  \tr_c \psekibun \frac{1}{(\epsilon_n - gA_4)^2 + E_p^2 }  
\end{equation}
does not depend on $\omega_n$ but the dispersive term  
\begin{eqnarray}
\Pi_\sigma^2  (\omega_n, q,  A_4) & = & \left( \omega_n^2 + \bq^2 + 4M_0^2 \right) F (\omega_n, \bq,  A_4 )  \\
\Pi_\pi^2 (\omega_n, q,  A_4) & = &  \left( \omega_n^2 + \bq^2 \right) F (\omega_n, \bq,  A_4 )
\end{eqnarray}
contains explicit $\omega_n$ dependence and  
\begin{eqnarray}
F (\omega_n, \bq ,  A_4) & = & 2 T \sum_{n'} \tr_c \psekibun  
\frac{1}{(\epsilon_{n'} - gA_4)^2 + E_p^2 }  
\nonumber \\
& & \qquad \times
\frac{1}{(\epsilon_{n'} - gA_4+ \omega_n)^2 + E_{p+q}^2 }  
\label{F}
\end{eqnarray}
where $\epsilon _n=(2n+1)\pi T$ are the quark Matsubara frequencies.

We now show that the contributions of the mesonic correlation to 
the pressure (\ref{Mpressure}) indeed contain
those of free meson gas composed of massless pions and massive sigma mesons.  
For this purpose we first eliminate the "tadpole" term in $\Pi_\sigma^1$ and $\Pi_\pi^1$ by using the stationary condition, or the "gap equation" (\ref{gap})
which determines the quark quasiparticle mass $M_0$ in the symmetry broken phase.
We then find, 
\begin{eqnarray}
(2 G)^{-1} -  \Pi_\sigma & = &   ( \omega_n^2 + \bq^2 + 4 M_0^2) F(\omega_n, q, A_4) 
+ {\displaystyle \frac{m_0}{2G M_0} } 
\nonumber \\
\label{smeson}\\
(2 G)^{-1} -  \Pi_\pi 
& = &   ( \omega_n^2 + \bq^2 ) F(\omega_n, q, A_4) + {\displaystyle \frac{m_0}{2G M_0} } 
\label{pmeson}
\end{eqnarray}

In the chiral limit, $m_0 = 0$, we can thus separate the contributions of the collective bare meson 
modes from non-collective individual excitations 
\begin{equation}
p_M (T, A_4) = p_M^{\rm coll.} (T) + p_M^{\rm non-coll.} (T, A_4) .
\end{equation}
The first term, independent of the color gauge potential, are identical to the pressures of free sigma meson 
gas with mass $m_\sigma = 2 M_0$ and of free massless pion gas with 3 isospin degeneracy.  

\begin{eqnarray}
p_M^{\rm coll.} (T) =  \int \frac{d^3 q}{(2\pi)^3}
\left[ \frac{q^2}{3\omega_q} f_B (\omega_q) + 3 \times \frac{q}{3}  f_B (q) \right] + p_M^0 
\nonumber \\
\end{eqnarray}
where $\omega_p = \sqrt{\bq^2 + 4 M_0^2}$ is the sigma meson energy and $f_B (\omega) = 1/( e^{\beta \omega} -1 )$ is a bosonic single particle
distribution function.   The mesonic vacuum pressure $p_M^0$ which depends on $M_0$  
will be removed at zero temperature by the renormalization, but we retain the deviation due to the shift of $M_0$ at finite temperature.  
The momentum integral for the pressure of the massless pion gas can be evaluated analytically and one finds a familiar Stephan-Boltzmann pressure  
with three-fold isospin degeracy: $p_{\rm pion}^{\rm free} = 3 \times \frac{\pi^2}{90} T^4 = 0.33 \times T^4$. 

The non-collective contribution 
depends on the color fields since it is composed of the underlying quark degrees of freedom.  
The sum over the bosonic sum over $\omega_n$ 
can be performed again by contour integration and we find, 
\begin{eqnarray}
p_M^{\rm non-col.}  
& =  & - 2 \int \frac{d ^3 q}{(2\pi)^3} \frac{1}{2\pi i}\int_0^\infty d \omega
\left[ 1 + \frac{2}{ e^{\beta \omega} -1} \right]  
\nonumber \\
& & \qquad \times 
\ln \left[ \frac{F( \omega + i \epsilon, q, A_4)}{F( \omega- i \epsilon, q, A_4)} \right]
\end{eqnarray}
where $F( \omega, q, A_4 ) $ consists of two terms
\begin{equation}
F(\omega, q, A_4 )  =   F_{\rm scat} (\omega, q, A_4 )  +
F_{\rm pair} (\omega, q, A_4 ) 
\end{equation}
The first term is the scattering term given by
 \begin{eqnarray}
F_{\rm scat} & = &\frac{1}{2} \int \frac{d ^3 p}{(2\pi)^3} \frac{1}{E_p E_{p+q}} 
 \frac{E_p - E_{p +q} }{ (E_p - E_{p +q})^2 -\omega^2 } 
\nonumber \\
& & \quad \times \tr_c \left( f (E_p -igA_4) - f (E_{p+q} - ig A_4 ) \right) 
\nonumber \\
\label{scattering}
\end{eqnarray}
while the second term corresponds to the pair creation and annihilation term 
 \begin{eqnarray}
F_{\rm pair} & = & \frac{1}{2}\int \frac{d ^3 p}{(2\pi)^3} \frac{1}{ E_p  E_{p+q}} 
\frac{E_p + E_{p +q} }{ (E_p + E_{p +q})^2 - \omega^2 }
\nonumber \\
& & \quad  \times \tr_c \left( 1 - f (E_p -igA_4) - f (E_{p+q} - ig A_4 ) \right) 
\nonumber \\
\label{pair}
\end{eqnarray}
Again, the external gauge fields appear as a phase factor in the quark distribution function as in the mean-field approximation.  
We therefore replace these phase factors by the Polyakov loops and then substitute them by statistical average 
as in (\ref{averagedis}):
\begin{eqnarray}
{\cal F} ( \omega, q ) & \equiv & \langle F( \omega, q,  A_4 ) \rangle  
= {\cal F}_{\rm scat} (\omega, q) + {\cal F}_{\rm pair} (\omega, q ) 
\nonumber
\end{eqnarray}
where  
$ {\cal F}_{\rm scat} ( \omega, q  ) $
and ${\cal F}_{\rm pair} (\omega, q ) $
are given by (\ref{scattering}) and (\ref{pair}), respectively, with the quark distribution replaced by (\ref{averagedis}).

The function 
${\cal F} (\omega \pm i \epsilon, q ) $ contains a real part ${\cal F}_1( \omega, q) $ and an imaginary part 
$ \pm {\cal F}_2 ( \omega, q ) $.  It is convenient to write 
\begin{eqnarray}
{\cal F} ( \omega \pm i \epsilon, q ) 
&=&  \sqrt{  {\cal F}_1^2 +  {\cal F}_2^2 } e^{\pm i \phi ( \omega, q) }  ,
\end{eqnarray}
where the argument $\phi$ is given by
\begin{equation}
 \phi (\omega, q) =  \tan^{-1} \frac{{\cal F}_2(\omega, q)}{{\cal F}_1(\omega, q)} ~~.
\end{equation}
The real part and imaginary part of the function ${\cal F}$ are further decomposed into two parts. 
The two components of the real part are given by the principal part of the integrals, 
while the two components of the imaginary part contains the energy conserving $\delta$-functions:
\begin{eqnarray}
{\cal F}_{\rm scat., 2}  
& = &  - \frac{\pi}{4}  \int \frac{d ^3 p}{(2\pi)^3} \frac{1}{E_p E_{p+q}}
\left( f_\Phi (E_p) - f_\Phi (E_{p+q} ) \right)
\nonumber \\
& & \qquad \times \left( \delta (\omega + E_p - E_{p +q}) - \delta ( \omega - E_p + E_{p +q} ) \right) 
 \nonumber \\ 
{\cal F}_{\rm pair, 2} 
& = &  - \frac{\pi}{4}  \int \frac{d ^3 p}{(2\pi)^3} \frac{1}{E_p E_{p+q}} 
\left( 1 - f_\Phi (E_p ) - f_\Phi (E_{p+q} ) \right) 
\nonumber \\
& & \qquad \times \left( \delta (\omega + E_p + E_{p +q}) - \delta ( \omega - E_p - E_{p +q} ) \right) 
\nonumber 
\end{eqnarray}
It is evident that the scattering term has a non-zero imaginary part in the space-like energy-momentum region 
($\bq^2 > \omega^2$), while the pair creation/annihilation term is non-vanishing only in the time-like region
($\bq^2 < \omega^2$).  
It is important to note that non-collective mesonic correlation arises only from non-vanishing 
imaginary part of ${\cal F} (q, \omega)$.
The pressure arising from the non-collective or individual excitations of the system is given by 
\begin{eqnarray}
p_M^{\rm non-col.} (T) & = &
 - 4 \int \frac{d ^3 q}{(2\pi)^3} \int_0^\infty \frac{d \omega}{2\pi}
\frac{1}{ e^{\beta \omega} -1} 2 \phi (\omega, q)
\nonumber \\
& & \qquad + \Delta p_M^{\rm non-col.}  
\end{eqnarray}
where the temperature independent background pressure $\Delta p_M^{\rm non-col.} $ contains the vacuum 
pressure due to the fluctuation of the pair of quark-triad and anti-quark triad which is removed again 
when we estimate the physical pressure at finite temperature.   

We note that foregoing analysis only applies to the chiral limit ($m_0 = 0$) and symmetry broken phase.
For non-zero value of $m_0$, the separation of the collective mode and the individual excitation is not 
as simple as above analysis due to the term $m_0/(2G M_0)$ in the dispersions (\ref{smeson}) 
and (\ref{pmeson}) of each kind of mesons. 
Also at high temperatures we have no broken-symmetry solution in the gap equation.      
In such cases we have to go back to the original expression (\ref{Mpressure}) in order to compute the 
mesonic correlation energy as has been done in \cite{RHRW08}. 
Collective meson poles appear separated from the continuum of individual excitations of a pair of 
quark triad and anti-quark triad and the former contribution dominates the pressure at low temperatures,
yielding meson gas equation of state again.
More detail discussion of this analysis is reported in \cite{Yamazaki:2012ux}.

We show in Fig. 2 the pressure calculated by the present method.  The upper panel (a) is the result computed in the 
chiral limit $m_0 = 0$.  At low temperatures below the second order chiral transition temperature $T_c$, 
the pressure from quark quasiparticle excitations are suppressed by the quenching of the distribution function:
only excitations of triad of quarks with color singlet configuration are allowed with effective excitation energy 
three times that of massive quark excitation, this contribution to the pressure is hence suppressed. 
The dominant mode of excitations which determine the pressure is due to massless pion excitation with 
3 isospin degeneracy as shown in the red solid curve.  
We found the contribution from massive sigma meson excitations and non-collective individual excitations of 
the quark triads and antiquark triads are also negligible at low temperatures.    
At high temperatures, massless quark excitations in the mean-field approximation dominate the pressure 
with comparable contributions of the gluon pressure phenomenologically introduced in the construction of 
the effective potential. 
Contribution from the mesonic correlations becomes negligible at high temperatures, and the pressure 
approaches to that of ideal massless quarks and gluons in our model. 
The deviation is due to the slight deviation of $\Phi$ from unity.  
With symmetry breaking finite bare quark mass, shown in the lower panel (b), the pion becomes massive and 
its pressure decreases exponentially at very low temperature, otherwise the behavior of pressure 
at higher temperatures is qualitatively unchanged.   

We have shown that an equation of state of meson gas can be derived from interacting quarks 
with a Nambu-Jona-Lasinio type interaction in uniform background color field. 
In the low temperature confining phase, the unwanted quark quasiparticle excitations in the mean-field 
approximation are removed by Fukushima's method, introducing the quenched quark distribution in the 
vanishing Polyakov-loop expectation value, which converts the distribution of individual quarks into that of 
triads of quark in color singlet configuration. 
The mesonic excitations which appear in the correlation energy as a correction to the mean-field approximation,
dominates in the low temperature confining phase.  
We also found that the corrections due to non-collective individual excitation of triads of quarks or antiquarks
are suppressed compared to collective meson excitations. 
Although each of these quark triads somewhat resembles a baryon carrying unit baryon number; these quarks carry the same amount of energy-momentum, hence not correlated spatially. 
It remains to be seen how these quark-triads are converted to real baryons taking into account internal
correlations among three quarks. 
For extensions of the present method to systems at finite baryon chemical potentials, such more realistic treatment 
of baryonic correlations would therefore be very important. 

We would like to thank H. Fujii for helpful discussions. We also thank G. Baym, J.P. Blaizot, T. Hatsuda, F. Lenz, B. Svetitsky for their interests in this work and comments. 
KY's work has been partially supported by the Global COE Program "the Physical Sciences Frontier", MEXT, Japan.   
TM's work has been supported by the Grant-in-Aid \# 21540257
of MEXT, Japan.

\includegraphics[clip,width=70mm]{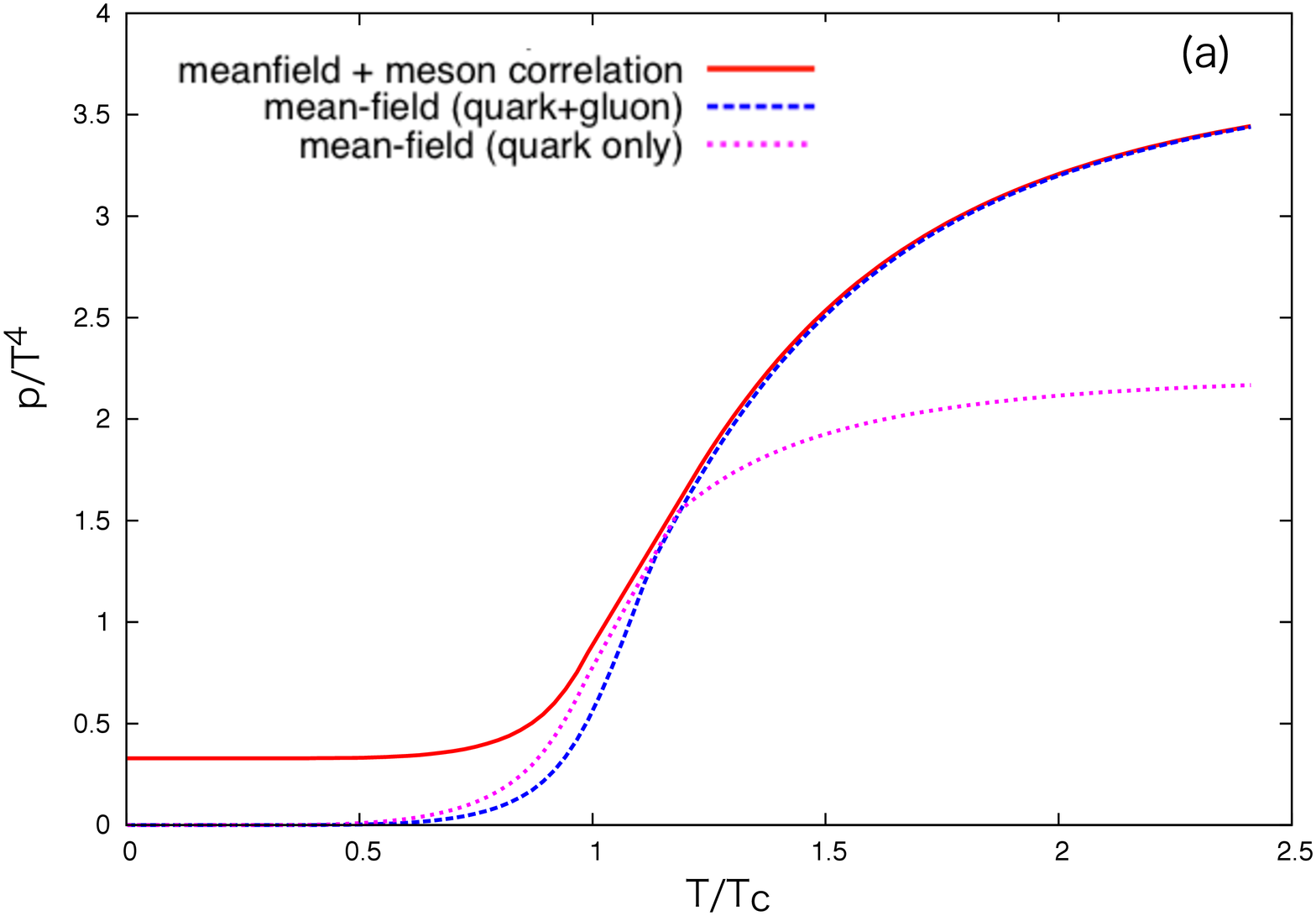}
\includegraphics[clip,width=70mm]{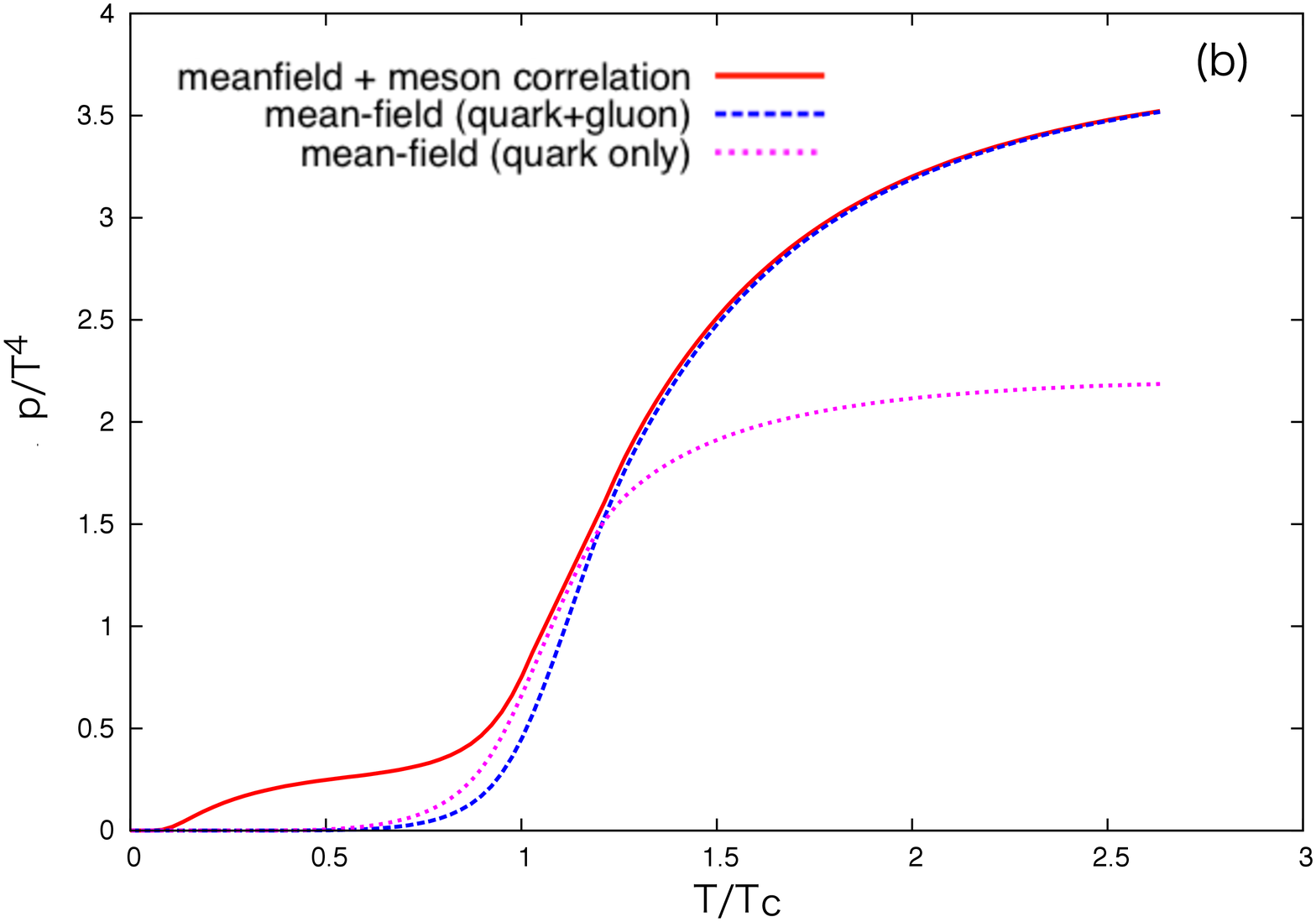}
\vspace{-5truemm}

\makeatletter
\def\tbcaption{\def\@captype{table}\caption}
\def\figcaption{\def\@captype{figure}\caption}
\makeatother
\figcaption{
Pressure scaled by $T^4$ as a function of temperature: 
the upper panel (a) is for vanishing bare quark mass (the chiral limit) and the lower panel (b) is with finite bare quark mass $m_0$ which is chosen to reproduce the pion mass $m_\pi = 140$MeV.  
In comparison, the pressures calculated in the mean-field approximation are shown with (without) gluon 
contributions. }

\newpage

\end{document}